\documentclass[useAMS,usenatbib,fleqn] {mnras}
\usepackage{epsfig}
\usepackage{amsmath}
\usepackage{graphicx}
\usepackage{amssymb}
\usepackage{pdflscape}

\title[Collisional nature of 2nd-generation stars in GCs]{Stellar collisions in globular clusters: the origin of multiple  stellar populations}

\author[Kravtsov et al.]{Valery Kravtsov$^{1}$\thanks{E-mail, VK: vkravtsov1958@gmail.com}, Sami Dib$^{2}$, Francisco A. Calder\'{o}n$^{3}$, Jos\'{e} Antonio Belinch\'{o}n$^{4}$\\
$^{1}$Sternberg Astronomical Institute, Lomonosov Moscow State University, University Avenue 13, 119899 Moscow, Russia\\
$^{2}$Max Planck Institute for Astronomy, K\"{o}nigstuhl 17, D-69117, Heidelberg, Germany\\
$^{3}$Departamento de F\'{i}sica, Universidad Cat\'{o}lica del Norte, Av. Angamos 0610, Antofagasta, Chile\\
$^{4}$Departamento de Matem\'{a}ticas, Universidad de Atacama, Av. Copayapu 485, Copiap\'{o}, Chile\\
}

\begin{document}
\maketitle

\date{Accepted 2022 March 10. Received 2022 February 22; in original form 2021 July 2}

\pagerange{\pageref{firstpage}--\pageref{lastpage}}
\pubyear{2016}
\label{firstpage}

\begin{abstract}

Two generations of stars, G1 and G2, typically populate Galactic globular clusters (GCs). The origin of G2 stars is unclear. We uncover two empirical dependencies between GC characteristics, which can be explained by the formation of G2 Main-Sequence (MS) stars due to collision\slash merging of their primordial counterparts (G1). A similar genesis of both G2 stars and peculiar objects like LMXBs and millisecond pulsars is also implied. Indeed, we find a significant (at a confidence level $> 99,9\%$) anti-correlation between the fraction of G1 red giants ($N_{G1}/N_{tot}$) and stellar encounter rates among 51 GCs. Moreover, a Milky Way-like initial mass function (IMF) requires at least $\sim50$\% of MS stars located in the mass range $[0.1-0.5] M_{\sun}$. Unlike cluster mass loss, stellar collisions\slash merging retain these G1 stars by converting them into more massive G2 ones, with mainly $M_{MS} > 0.5 M_{\sun}$. This process coupled with a decreasing relative mass loss with increasing GC masses implies a smaller ($N_{G1}/N_{tot}$) in more massive GCs with a shallower present day MF. From data for 35 GCs, we find that such an anti-correlation is significant at 98.3\% confidence level (Spearman's correlation) for the 12 most massive GCs ($M_{GC} > 10^{5.3} M_{\sun}$) and it is at a confidence level of 89\% for the 12 least massive GCs ($M_{GC} < 10^{5.1} M_{\sun}$). Other fractions of G1 and G2 stars observed at the bottom of the MS as compared with the red giant branch in a few GCs are consistent with the scenario proposed.

\end{abstract}

\begin{keywords}
globular clusters: general - (stars:) binaries (including multiple): close
\end{keywords}

\section{INTRODUCTION}\label{introduc}

Globular clusters (GCs) are amongst the most ancient stellar aggregates formed both in the Milky Way and in other galaxies. They form very early during the galaxies' assembly history and despite a continuous evaporation of stars, have remained bound in different galactic environments. Thanks to their unique characteristics, GCs are of special interest for studying a variety of key problems in astrophysics.

One well known problem arises from the observational fact that stellar populations (SPs) in GCs are not quite simple. A lot of evidence about a more complex nature of GC SPs have been obtained in numerous works up to now. This complexity is widely referred to as the multiplicity of the cluster SPs. Here, we are focusing on the most generic case of the multiplicity observed in virtually all but several monometallic Galactic GCs. Namely, two populations or generations of stars, G1 (primordial) and G2 (secondary), with no (or no significant) difference in mean iron abundance are typically distinguished both spectroscopically \citep[e.g.,][]{Carretta2009} and photometrically \citep{Milone_etal2017} in GCs. Hereafter, we refer to this multiplicity as a complexity of SPs. Moreover, the complexity of SPs has been studied in a wider context, i.e not only in Galactic GCs but also in young and intermediate-age massive star clusters in the Large and Small Magellanic Clouds (LMC and SMC). These clusters normally exhibit a complexity of their SPs at the age around 2 Gyr in the form of an extended MS turnoff \citep{Mackey_BrobyNiel2007,Mackeyetal2008,Milonetal2009}. The manifestations typical for multiple SPs in GCs are observed in LMC/SMC massive star clusters older than 2 Gyr \citep[e.g.,][]{Hollyheadetal2017,Niederhoferetal2017,Hollyheadetal2018,Milonetal2020,Lietal2021}. This has challenged our understanding of the origin of the second generation in both massive star clusters in general and GCs in particular.

A number of scenarios have been proposed already a decade ago and even earlier to explain both the origin of and key differences between the SPs in GCs \citep[e.g.,][]{Decressin_etal2007,Ventura_DAntona2009,Decressin_etal2010,Valcarce_Catelan2011,Krausetal2013}. For the past decade, many aspects of this important subject were further studied in more detail. They have been summarized and reviewed, in particular, by \citet{BastianLardo2018}. Interested readers are referred to this paper for key particulars concerning the subject, including the scenarios proposed so far and problems met within the frame of each one. It appears that, although a much deeper insight into the matter has been gained, the origin of G2 stars remains an unclear and even more controversial issue than before. One of the most serious problems associate with each of the proposed scenarios is the so-called "mass-budget problem". Here, we report supporting evidence of essential contribution of stellar collision\slash merging to the formation of G2 stars, a mechanism that is probably able not only to overcome the "mass-budget problem" but to simultaneously explain other features of massive clusters such as the origin of their initial mass function (IMF) \citep[e.g.,][]{Dibetal2007}.

We have recently paid attention to modified MS stars which should have formed from their primordial counterparts due to (presumably mainly binary mediated) stellar collisions in the central regions of GCs \citep{Kravtsov2019,Kravtsov2020,Kravtsov_Calderon2021}. In particular, we argued for a potential role of these stars and their descendants on both the sub-giant branch (SGB) and red giant branch (RGB) in contributing toward creating (1) a range of masses and varying surface abundances of particular elements among stars naturally believed to be congenerical (i.e., which evolve(d) from primordial MS stars of the same mass with its negligible dispersion) and (2) radial effects among MS turnoff and SGB/RGB stars in GCs. Indeed, a small portion of such modified stars are well known as collisional blue stragglers (CBSs) actually observed in GCs \citep{Ferraro2009,Dalessandro2013}. The existence of CBSs is direct evidence of the collision process at work. However, there are reasons to believe that the amount of the presently observed CBSs is only "the tip of the iceberg" of the total number of modified MS stars formed and accumulated in GCs during their lifetime, in particular within the mass range $[0.5-0.9] M_{\sun}$. By their characteristics, these collision-generated MS stars are different from primordial MS stars of the same mass and could hardly therefore belong to the population of G1 stars in GCs. For instance, according to the models of collision products appropriate to an old open cluster like M67 \citep{Glebbeek_etal2008}, their descendants on the RGB are expected to have different surface carbon and nitrogen abundances in comparison with normal red giants in the sense that [C/Fe] is depleted and [N/Fe] is enhanced (see also \citet{Weietal2020} who calculated surface chemical anomalies of low-mass MS stars due to mass transfer in a binary system). The so-called luminous red novae (see Section~\ref{verif_eff}) are known to be reliable evidence of merging binary systems in over a large mass range closely approaching the most typical mass of binary systems in GCs and are promising to shed more light on the role and outcome of the merging process. Moreover, collisionally formed MS stars normally complete their MS evolution and achieve (and go on) the RGB simultaneously with primordial stars of lower mass. This can be not only suspected based on basic laws of stellar astrophysics but also more reliably deduced from relevant models of MS stars formed due to collision of two primordial MS stars with different mass ratios in GCs, at different cluster age \citep{Sills_etal2009}. However, one of the main problems is that the fraction of collisionally formed G2 stars in GCs is poorly known. It is estimated with a large uncertainty.

In this paper, we further study the role and effect of the collision process, binary-mediated and permanently acting in GCs, in the transformation of their stellar populations, in general. We explore the dependencies, specifically the fraction of G1 stars versus both the stellar encounter rates and mass function (MF) slope in GCs, giving insight into a non-negligible mechanism for G2 stars to form.

In the next section (Section~\ref{analysis}), we describe the data used for our analysis. In Section~\ref{results}, we present: (1) the results dealing with both new dependencies obtained from the data used; (2) a model testing the availability of G1 MS stars in initial GCs to form populations of G2 stars observed in the present-day GCs on the RGB; (3) a verifiable effect implied by our scenario and observational results on merging stars producing luminous red novae. A summary of our results and concluding remarks are in Section~\ref{conclusions}.

\section{DATA} \label{analysis}

For the present study devoted to the contribution of stellar collisions to the formation of the populations of G2 stars in GCs, we used three key sets of recently published and publicly available data on Galactic GCs and their SPs. First, we use the atlas on multiple SPs in Galactic GCs, obtained by \citet{Milone_etal2017}. Using uniform multiband HST photometry \citep{Sarajedini_etal2007,Piotto_etal2015} in the central parts of a large sample of 57 GCs, the authors isolated RGB stars, separated them into subpopulations belonging to different generations, and deduced a number of parameters characterizing each generation. Also, \citet{Milone_etal2017} calculated the total number of RGB stars, $N_{tot}$, falling in the observed area of each GC and quantified the relative number of G1 red giants ($N_{G1}/N_{tot}$) as the ratio between the number of G1 ($N_{G1}$) and the total number of red giants in a GC. The number of GCs where \citet{Milone_etal2017} have been able to estimate the relative number of G1 red giants is 54.

Second, we also make use of the recently published data of \cite{Ebrahimietal2020} on the clusters' stellar MFs, masses, and a number of other characteristics listed for a sample of 32 GCs. However, NGC 6584 was excluded from our analysis, since the data on this GC are incomplete. Instead, we added data, taken from \citet{Sollima_Baumgardt2017} on four other GCs: NGC4833, NGC6205, NGC6397, and NGC6656. Therefore, we finally obtained a samle of 35 GCs with available data on both the fraction of G1 and MFs and cluster masses. The original papers list logarithms of both the luminous and dynamical masses of GCs, log$M_{lum}$ and log$M_{dyn}$, respectively. We notice that the GC masses used for our present study are expressed in the form $M_{GC} = 10^{A} M_{\sun}$, where the exponent $A$ is the mean between $A_{lum}=$log$M_{lum}$ and $A_{dyn}=$log$M_{dyn}$.

Third, the data on stellar encounter rates (and their errors) in GCs, $\Gamma$, are taken from \citet{Bahrametal2013} where the data are normalized assuming $\Gamma = 1000$ for the GC NGC104. Since the values of log$\Gamma$ were used instead of $\Gamma$, we calculated these quantities and expressed the corresponding errors in the logarithmic scale. Overall, we obtained a sample of 51 GCs with available data on both $\Gamma$ and ($N_{G1}/N_{tot}$). Finally, we refer the interested readers to the aforementioned original publications, for additional detail on both the data themselves and related information about the observations and data reduction procedures.

\section{RESULTS}\label{results}

We examine the potential key role of binary-mediated stellar collisions in the formation of multiple SPs in GCs through the formation of modified MS stars (a small portion of which are presently observed as collisional BSs) from primordial MS stars, at different epochs. Many observational studies show that the typical effects of multiplicity of SPs in GCs, such as anti-correlations between abundances of some chemical elements, are not manifested by SPs even in old Galactic open clusters nor in the field, except perhaps for its dense regions in the Galactic bulge \citep{Schiavonetal2017}. This fact implies that one of the most probable factors responsible for this dissimilarity is binary-mediated stellar collisions which do normally occur in the densest parts of GCs but are much less probable to occur in environments where the stellar density is low. In relation to this, it is relevant to mention that a new type of somewhat exotic objects (at least due to their special location in the color-magnitude diagram of globular and open clusters), so-called sub-subgiant stars, was recently introduced and studied by \citet{Gelleretal2017a,Gelleretal2017b} and \citet{Leineretal2017}. These authors argue that a portion of these stars are the product of merged MS stars, which are now "observed while settling back down onto the normal main sequence".
Given the aforementioned arguments, we verify two expected outcomes of the process of collision-induced merging of stars in relationship with the multiplicity of SPs in GCs.

\subsection{The fraction of G1 stars versus the stellar encounter rates} \label{gamma_g1}

Extremely dense central regions of GCs are recognized as unique sites where exotic objects, like collisional BSs, millisecond pulsars, and low mass X-ray binaries (LMXBs), among others, are preferably observed. It is primarily due to the conditions favorable for the formation of close binaries, subsequent evolution of which leads to their transformation into such exotic objects responsible for the respective astrophysical phenomena. Binary stars play a crucial role in relation to all processes that occur in the central regions of GCs, in particular, and in the dynamical evolution of GCs, in general. This important topic and many closely related aspects like binary mediated collisions and their outcomes were the subject of numerous astrophysical investigations. For many details and respective key references, we refer here, for example, to the review by \citet{Meyl_Hegg1997}. As has been already noted, we assumed that a significant fraction of G2 stars, like their prototypes in the form of collisional BSs, are of binary mediated collisional origin.

\begin{figure}
\includegraphics[angle=-90,width=\columnwidth]{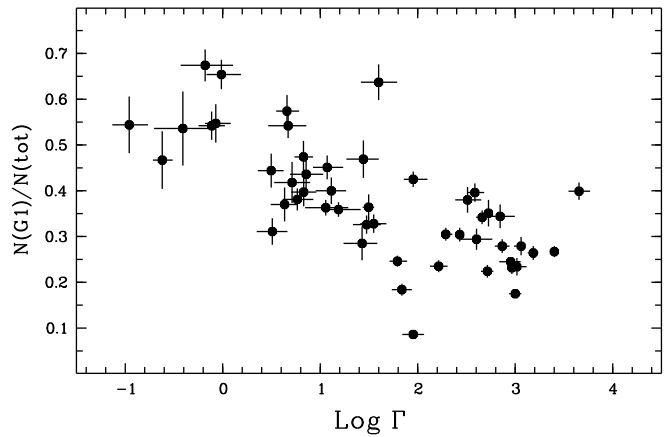}
\caption{Dependence of the fraction of G1 stars on logarithm of the stellar encounter rate among 51 Galactic GCs. See the paper text for more detail. $\Gamma$ are calculated and normalized assuming $\Gamma = 1000$ for the GC NGC104 by \citet{Bahrametal2013}.}
\label{fig:fig1}
\end{figure}

It is now established that there is statistically significant correlation between the populations of both LMXBs and millisecond pulsars and the stellar encounter rate in Galactic GCs \citep{Pooleyetal2003,Bahrametal2013}. Given these facts, we explored the dependence of ($N_{G1}/N_{tot}$) on log$\Gamma$, the logarithm of the stellar encounter rate in GCs, by relying on the respective data available for a sample of 51 Galactic GCs. This is displayed in Figure~\ref{fig:fig1} which indicates an anti-correlation between the two quantities. Spearman's correlation coefficient is $\rho = -0.729$ and this implies that this anti-correlation is statistically significant at a high confidence level ($> 99,9\%$). We note that the GC that deviates most from the sequence shown in Figure~\ref{fig:fig1} and has the smallest ($N_{G1}/N_{tot}$) is Omega Cen. It has not been excluded from the statistical test made. However, this GC is obviously a special case among Galactic GCs, since it is well known to show a fairly large star-to-star variation in the [Fe/H] ratio and the presence of genuine multiple stellar populations appropriate for galaxies. This is opposed to the complexity of stellar populations in the majority of Galactic GCs, which we deal with in the present study. So, the astrophysical concept of the parameter ($N_{G1}/N_{tot}$) in Omega Cen is not, strictly speaking, equivalent to that in other GCs. There is also another important point concerning the ($N_{G1}/N_{tot}$) ratio that should be commented on. It has been measured in the central parts of GCs. Hence, it characterizes the fraction of G1 stars (red giants) exactly in the central part of a GC but not in the GC as a whole. The measured values of this parameter should not be considered as absolute ones, i.e. appropriate for GCs as entire objects. Indeed, given that G2 stars are often more concentrated toward the GC centers, the expected absolute values of the fraction of G1 stars in GCs are expected to be, on average, systematically higher than the presently available data on the ($N_{G1}/N_{tot}$) ratio used in the present study.

Finally, it should be noted that the obtained dependence is not quite trivial, for different reasons. It is seen that the dispersion of the points is obviously larger than would be expected being due to the typical formal errors of both quantities and the deviations of two GCs (apart from Omega Cen) are notably large. We will analyze this dependence in more detail in a forthcoming paper.

\subsection{The fraction of G1 stars vs the slope of the present-day mass function} \label{mf_g1}

There is ongoing, intense debates on whether the IMF of stars in individual stellar clusters is universal and similar to the mass function of stars in the nearby Galactic field \citep[e.g.,][]{Bastianetal2010,Dib2014,Dibetal2017}. If the IMF of GCs is assumed to be canonical, similar to the MF of stars in the Galactic field irrespective of GC characteristics, then the bulk of MS stars ($\sim85$\%) should have masses $M_{MS} < 0.5 M_{\sun}$ or more specifically $\sim50$\% with $[0.1-0.5] M_{\sun}$ \citep{Kroupa2001,Kroupa2002}. The fraction of these low-mass stars in GCs decreases with GC age. One of the important processes responsible for that is dynamical interactions. At first glance, stellar collisions leading to merging of stars produce a similar effect, since they primarily reduce this reservoir of the most numerous stars. However, in contrast to cluster evaporation, stellar collisions transform these low-mass G1 stars and thereby      retain them in GCs in other form, by converting them into G2 stars of a higher mass, primarily within the mass range $[0.5-0.9] M_{\sun}$. Therefore, the collisional formation of G2 stars implies that the shallower the actual stellar MF of GCs at a given cluster mass, the smaller the fraction of G1 stars. Recent simulations of dynamical cluster disruption \citep{Reina_Cam_etal18} predict decreasing relative mass loss with increasing GC masses. Given this, one can expect that the effect of stellar collisions on the transformation of the IMF should be more competitive among the most massive GCs.

In the collision-based scenario for the formation of G2 stars, an increasingly shallower present-day mass function would imply a more efficient collision process under the assumption of the same IMF for G1 stars. This would result in an increasingly smaller fraction of G1 stars and larger fraction of G2 stars. We will discuss this from a theoretical point of view in Section~\ref{toy_mod}. This process can also be impacted by variations in the IMF of G1 stars, particularly variations that could affect the initial fraction of stars present in the mass range $[0.1-0.5] M_{\sun}$. We will explore the effect of IMF variations in a forthcoming paper.

\begin{figure}
\includegraphics[angle=-90,width=\columnwidth]{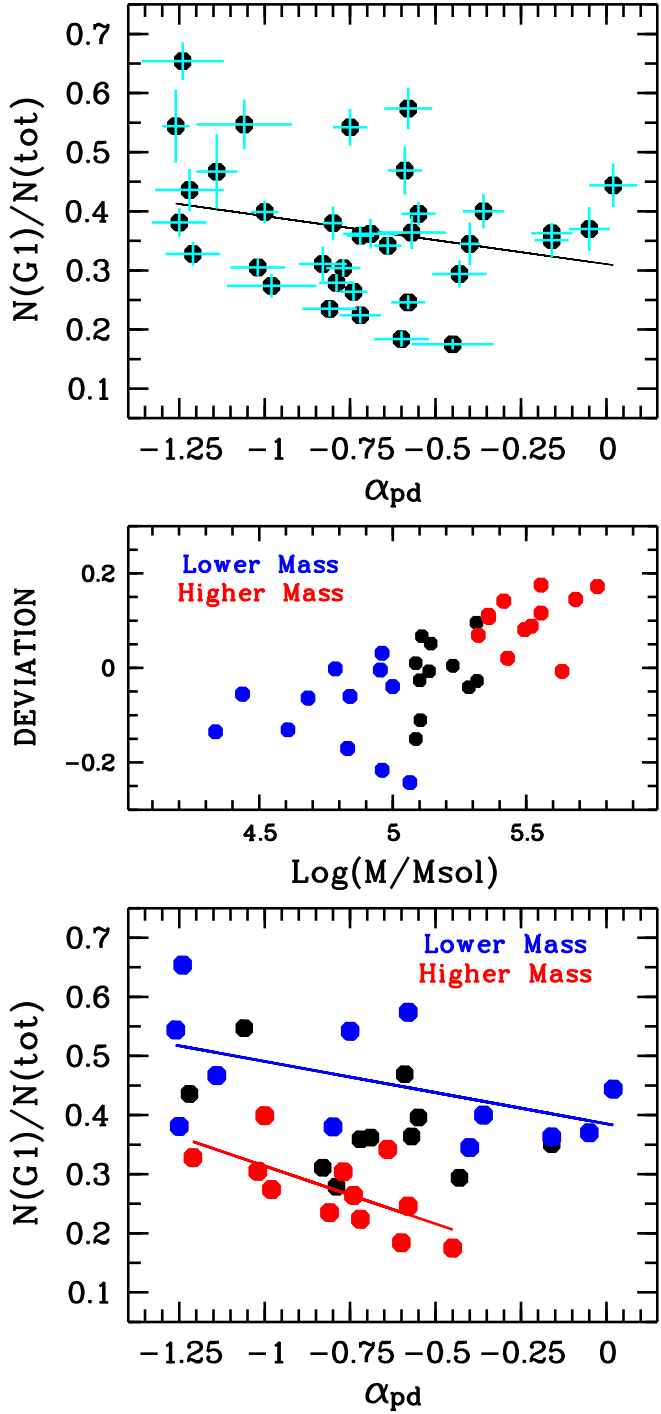}
\caption{Dependence of $N_{G1}/N_{tot}$, the fraction of G1 (first generation) stars, on power-law MF slope $\alpha_{pd}$ characterizing the present-day MF of Galactic globular clusters. Upper panel: general dependence for all the 35 Galactic globular clusters with available data; error bars are shown with cyan lines and the continuous line is a linear fit to the data. Middle panel: the deviations of individual points from the line fitted to the data as a function of the GC mass;
Lower panel: the same as in the upper panel, but the dependence is shown for three subsamples of GCs separated by mass. The red and blue dots are 12 GCs of higher mass ( $M_{GC} > 10^{5.3} M_{\sun}$) and 12 GCs of lower mass ($M_{GC} < 10^{5.1} M_{\sun}$), respectively, while 11 GCs with intermediate mass are shown by the black dots.}
\label{fig:fig2}
\end{figure}

We derived the dependence between the present-day stellar MF slope and the fraction of G1 stars in GCs of a large range of mass. This is shown in the upper panel of Figure~\ref{fig:fig2}. It appears quite dispersed, with correlation between the two quantities at low confidence level of 70\% according to Spearman's rank correlation test. We have drawn a linear fit to the data, obtained the deviations of individual points from the line fitted to the data, and plotted them as a function of the GC mass in the middle panel of Figure~\ref{fig:fig2}. From this panel, one can see that the deviations depend on the mass of GCs. For this reason, the analyzed sample of GCs has been divided into subsamples by meeting the following requirements: as large as possible separation by mass between higher- and lower-mass GCs along with having a reasonable subsample size of objects in each category. We decided upon a minimum subsample size of 12 GCs in each category. High-mass GCs have masses $M_{GC} > 10^{5.3} M_{\sun}$ and lower-mass GCs have masses $M_{GC} < 10^{5.1} M_{\sun}$ The remaining 11 GCs fall in the intermediate mass range. GCs belonging to the three subsamples are shown by different colors in both the middle and lower panels of Figure~\ref{fig:fig2}. Spearman's rank correlation test shows that a correlation in the expected sense is statistically significant at 98.3\% confidence level for the subsample of the most massive GCs. For the subsample of the least massive GCs, however, there is more scatter, and the correlation is less significant with a confidence level of 89\% (and at least formally it is statistically insignificant). See test results in Table~\ref{tab:table_1}. From the appearance of the dependence for GCs of different mass in the lower panel of Figure~\ref{fig:fig2}, one cannot exclude potentially interesting details. For example, the range of the MF slope of the least massive GCs is obviously larger than among the most massive ones. Perhaps this is a signature of a larger mass loss from the least massive GCs and its respective stronger impact on the observed variation of the MF slope.

\begin{table}
	\caption{Spearman's rank correlation test.}
	\label{tab:table_1}
\begin{tabular}{clll}
		\hline
Mass range & $N_{GC}$ & $\rho^a$ & P(\%)$^b$ \\
		\hline
Total & 35 & -0.172 & 70.0\\
$M_{GC} > 10^{5.3} M_{\sun}$ & 12 & -0.664 & 98.3\\
$M_{GC} < 10^{5.1} M_{\sun}$ & 12 & -0.483 & 89.0\\
		\hline
\end{tabular}\\

\footnotetext{a}{$^a$ Spearman's $\rho$ coefficient of anti-correlation\\ between the MF slope $\alpha_{pd}$ and ($N_{G1}/N_{tot}$).}\\
\footnotetext{b}{$^b$ The probability of non-random deviation of \\ the $\rho$ coefficient from zero.}

\end{table}

We recall that understanding the origin of the complexity of stellar populations in GCs is mainly challenged by two key problems to be explained together, namely the "mass-budget problem" and the anti-correlations of the surface abundances of a number of light elements. Below, we deal with both issues in more detail.

\subsection{Mass budget: a toy model} \label{toy_mod}

Here, we present a model that can help us interpret our finding for the correlation between the ratio ($N_{1}(G1)/N_{tot}$) and the slope of the present-day MF $\alpha_{pd}$ in the framework of our proposed collision-based scenario. Strictly speaking, understanding the exact effect of collisions is obviously a very complicated task. It requires a reliable knowledge of a number of key parameters such as the collision rates over the entire stellar mass range, the coalescence efficiency as a function of stellar mass \citep[e.g.,][]{Dibetal2007} and the mass loss as a function of cluster mass, etc. However, in order to estimate whether, and under which conditions, the capacity of the main reservoir of primordial lower-mass MS stars would be sufficient to produce the amount of collision products comparable with the observed fraction of G1 stars in GCs, we apply a much simpler approach.
By means of our model which assumes a Kroupa-like IMF for the G1 stars, we compare the fractions of the MS stars that collide to form the respective G2 stars of a higher mass. Moreover, a model MS analogue of the ratio ($N_{1}(G1)/N_{tot}$) (empirically derived for the RGB in Galactic GCs by \citet{Milone_etal2017}) and the slope of the present day MF ($\alpha_{pd}$) are deduced to obtain a model MS analog of the empirical dependence obtained for the RGB and presented in Figure~\ref{fig:fig2}.

For simplicity, (1) we are focusing on the mass range $[0.1-0.9] M_{\sun}$ in which the bulk of MS stars reside in both young and present-day GCs, (2) we divide it by two ranges, assuming that stars that merge are those exclusively found in the mass range $[0.1-0.5] M_{\sun}$, and (3) we consider the case of insignificant mass loss, i.e. more compatible with the most massive GCs. The former mass range is conditionally referred to as "source" mass range, since it is very probably the main reservoir of stars. The majority of the respective collision products will fall in the next mass range $[0.5-0.9] M_{\sun}$ referred hereafter to as "target" mass range. Recall that the actually observed cluster RGB stars, which have been used (in the centers of GCs) by \citet{Milone_etal2017} to estimate the parameter ($N_{1}(G1)/N_{tot}$), are the descendants of former MS stars fallen in this mass range, with their mass $\sim0.8 M_{\sun}$ (primordial MS stars) or somewhat higher (collision products). Note that the requirement about collisions between G1 stars exclusively belonging to the source mass range is a too strong constraint, since these G1 stars might also collide with G1 stars from the target mass range by forming G2 stars with mass that falls within the target mass range. In the framework of our scenario, this latter collision channel would be more favorable for the formation of G2 stars from the point of view of resolving the mass-budget problem. Indeed, this channel implies that only one G1 star from the source mass range would be consumed to contribute one G2 star to the target mass range in contrast to our model requiring two G1 stars from the former mass range to form one G2 star for the latter range.

As an IMF for the G1 stars, we employ a Galactic field-like mass function \citep{Kroupa2001,Kroupa2002}. This is just a first approximation. It is evident that the real picture is more complicated and significantly nuanced. For convenience, we accept all the collisions in any initial GC as if they occurred at a time, sometime after the formation of primordial generation of stars G1.

If $N_{1}(G1)$ and $N_{2}(G1)$ are the total numbers of stars of the first generation that fall in the mass range $[0.1-0.5] M_{\sun}$ and $[0.5-0.9] M_{\sun}$, respectively, then the fractions of these two populations of stars to the initial total number of stars ($N'_{tot}$) can be written as:

\begin{equation}
f_{1,in}=\frac{N_{1}(G1)}{N'_{tot}},
\label{eq1}
\end{equation}

and

\begin{equation}
f_{2,in}=\frac{N_{2}(G1)}{N'_{tot}}
\label{eq2}
\end{equation}

If $N_{1,ext}(G1)=f_{ext} N_{1}(G1)$ is the total number of stars that have experienced a collision then the new total number of stars becomes:

\begin{equation}
N_{tot}=N'_{tot}-\frac{f_{ext} N_{1}(G1)}{2}
\label{eq3}
\end{equation}

After the collision process, a second generation of stars (i.e., G2 stars) is formed in the mass range $[0.5-0.9] M_{\sun}$. The new fractions of stars in the above-mentioned mass ranges will be given by:

\begin{equation}
\begin{array}{l}
f_{1,fl}=\frac{N_{1}(G1)-f_{ext}N_{1}(G1)}{N_{tot}}=\frac{N_{1}(G1)(1-f_{ext})}{N'_{tot}-f_{ext}N_{1}(G1)/2}=
\\
\\
=\frac{2 f_{1,in} (1-f_{ext})}{2-(f_{ext}f_{1,in})}
\end{array}
\label{eq4}
\end{equation}
and
\begin{equation}
\begin{array}{l}
f_{2,fl}=\frac{N_{2}(G1)+f_{ext}N_{1}(G1)/2}{N_{tot}}=\frac{N_{1}(G1)(f_{2,in}/f_{1,in}+f_{ext}/2)}{N'_{tot}-f_{ext}N_{1}(G1)/2}=
\\
\\
=\frac{2 f_{2,in}+f_{ext} f_{1,in}}{2-(f_{ext}f_{1,in})}
\end{array}
\label{eq5}
\end{equation}

The value of $g_{1}$ is defined for MS stars in the mass range $[0.5-0.9] M_{\sun}$ like its analog for RGB stars (i.e., $N_{1}(G1)/N_{tot}$) as:

\begin{equation}
\begin{array}{l}
g_{1}=\frac{N_{2}(G1)}{N_{2}(G1)+N_{2}(G2)}=\frac{N_{2} (G1)} {(N_{2}(G1)+f_{ext}N_{1}(G1)/2)}=
\\
\\
=\frac{2 f_{2,in}}{2 f_{2,in}+f_{ext} f_{1,in}}=\frac{1}{1+f_{ext} f_{1,in}/2 f_{2,in}}
\end{array}
\label{eq6}
\end{equation}

\begin{figure}
\includegraphics[angle=0,width=\columnwidth]{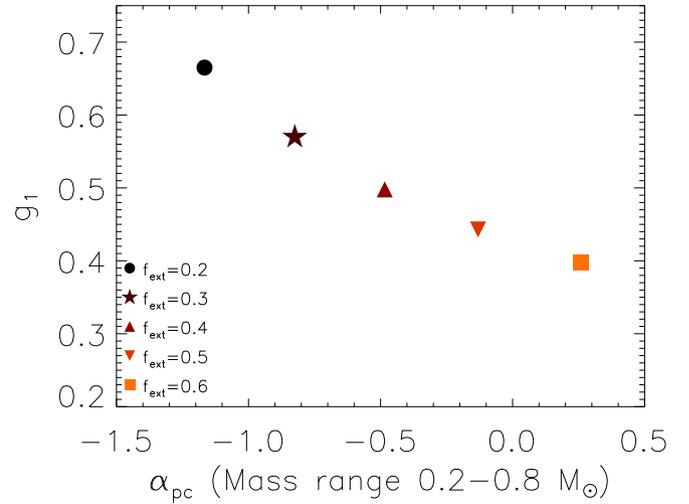}
\caption{Model relationship between the $g_{1}$ parameter, the MS analog of the ($N_{1}(G1)/N_{tot}$) parameter empirically derived for the RGB in GCs by \citet{Milone_etal2017}, and the slope $\alpha_{pc}$ (the model analog of of the present-day mass function $\alpha_{pd}$), corresponding to different values of $f_{ext}$ between 0.2 and 0.6. The canonical Kroupa IMF has been applied.}
\label{fig:fig3}
\end{figure}

Neglecting the effects of dynamical evolution during the collision process, the model values of the slope in the post collision MF ($\alpha_{pc}$) would be the result of the stellar collision process. For any functional form of the IMF of G1 stars, and for a given value of $f_{ext}$, we can estimate the values of $f_{1,in}$, $f_{2,in}$, which are used to derive the values of $f_{1,fl}$ and $f_{2,fl}$. With $f_{1,fl}$ and $f_{2,fl}$, we can then derive the values of $g_{1}$ and $\alpha_{pc}$. We can calculate $\alpha_{pc}$ as being

\begin{equation}
\alpha_{pc}=\frac{({\rm log}(f_{2,fl})-{\rm log}(f_{1,fl}))}{({\rm log}(0.7)-{\rm log}(0.3))},
\label{eq7}
\end{equation}

where 0.3 and 0.7 (in units of $M_{\sun}$) are the central values in the mass bins $[0.1,0.5] M_{\sun}$ and $[0.5,0.9] M_{\sun}$, respectively. We recall that the values of $\alpha_{pd}$ characterizing the present-day MF in the observational sample of GCs vary between $\approx 0.05$ and $\approx -1.25$ and these values have been derived by \citet{Ebrahimietal2020} assuming that the present-day MF of GCs is a single power law over the mass range $[0.2-0.8] M_{\sun}$. The total range considered for the two mass bins in the model is $[0.1-0.9] M_{\sun}$ and it encompasses the stellar mass range over which the value of $\alpha_{pd}$ is estimated in the observations. When calculating the fractions $f_{1,in}$ and $f_{2,in}$, and subsequently $f_{1,fl}$ and $f_{2,fl}$, we assume that the minimum stellar mass is $M_{min}=0.01$ M${\sun}$ and the maximum stellar mass is $M_{max}=150$ $M_{\sun}$. We build a series of models relying on a Kroupa-like IMF (i.e., the pre-collision IMF) and with various values of $f_{ext}$. For each model, we derive the corresponding values of $g_{1}$ and $\alpha_{pc}$. We show in Figure~\ref{fig:fig3} a set of models that tie the value of $g_{1}$ to $\alpha_{pc}$ for various values of $f_{ext}$ and which all employ the Kroupa IMF for the pre-collision G1 population. We can now compare it with the relation between the parameter ($N_{1}(G1)/N_{tot}$) and $\alpha_{pd}$ in Figure~\ref{fig:fig2}.

Figure~\ref{fig:fig3} shows that for the canonical Kroupa IMF, the capacity of the source mass range (i.e., the initial fraction of primordial MS stars formed in this mass range) would be sufficient to form large fractions of G2 MS stars, depending on the $f_{ext}$ value, comparable to those empirically derived on the RGB in many GCs. In other words, the model fraction $g_{1}$ of G1 MS stars left in the target mass range $[0.5-0.9] M_{\sun}$ after collisions have occurred, follows the same trend in terms of its dependence on $\alpha_{pd}$ as the one seen in the observations, in general agreement with the empirically deduced parameter ($N_{1}(G1)/N_{tot}$). The parameter $f_{ext}$ can formally run from 0 to 1 and result in lower and lower values of the fraction $g_{1}$ that would be more and more approaching to the minimum values of the parameter ($N_{1}(G1)/N_{tot}$) in GCs. However, the real maximum value of $f_{ext}$ defining the maximum fraction of G1 stars from the source mass range to be converted into the maximum fraction of G2 stars (minimum $g_{1}$) in the target mass range, can hardly approach to 1. Our numerical experiments show that values of $f_{ext} > 0.6$ yield too large values of $\alpha_{pc}$. So, we accept $f_{ext} \sim 0.6$ as being around the limiting value in this respect. Coming back to Figure~\ref{fig:fig3}, the minimum value of the parameter $g_{1}$ (in the target mass range) for the limiting case of $f_{ext} = 0.6$ is about $g_{1} \approx 0.40$. $f_{ext} = 0.5$ would perhaps be more realistic value for the limiting case.

It should be taken into account that an additional fraction of G1 stars from the source mass range will be converted into G2 stars within the same mass range. The spirit of this simple model is that the fraction of G2 stars is expected to be (much) smaller than G1 stars in the source mass range, in contrast to the target mass range. This additionally implies that $g_{1}$ in this mass range would normally be higher (larger fraction of G1 stars) than $g_{1}$ in the next mass range,  $[0.5-0.9] M_{\sun}$, and therefore on the RGB, too. Let us consider, for simplicity, a limiting case where all the formed G2 stars are kept in the source mass range, $[0.1-0.5] M_{\sun}$. Then, given that two merging G1 stars result in a single G2 star, we immediately conclude that the fraction of G2 stars should remain less than the fraction of G1 stars as long as $f_{ext} < 2/3$. In reality, this is impossible, since at least half of the G2 stars fall in the target mass range. So, the fraction of G2 stars in the source mass range (especially at $M_{MS} < 0.2-0.3 M_{\sun}$, see next Subsection) should be smaller or even much smaller (at a given value of $f_{ext}$), than in this limiting case. Given the probable partition between maximum fractions of G1 stars from the source mass range, consumed for the formation of G2 stars in both mass ranges, we finally come to the total limiting fraction of G1 stars allowed to collide in the source mass range around 0.70. This corresponds to 35\% of the total number of stars in an initial GC at these most favorable conditions.

As we noted at the beginning of the given Subsection, the above estimate of $g_{1}$ is an approximation. In principle, we should apply a correction reflecting the obvious fact that a portion of G1 stars from the target mass range, $[0.5-0.9] M_{\sun}$, will participate in collisions and form G2 stars. Therefore the fraction $f_{2,in}$ will effectively be smaller, resulting in a lower $g_{1}$. Instead of $f_{2,in}$ in Equation~\ref{eq6}, we then use $f_{2,in}(1-f_{ext}g_{1})$ where the factor $(1-f_{ext}g_{1})$ softly takes the above-mentioned fact into accounts. It would be natural to apply a stronger factor $(1-f_{ext})$ if the reducing population of G1 stars in this mass range was not simultaneously blended with increasing population of G2 stars mostly originated from the source mass range. So, in order to be more conservative and realistic, we multiply $f_{ext}$ by the first approximation value of $g_{1}$. The corrected value of $g_{1}$ will finally be around $g_{1} \approx 0.32$ for $f_{ext} = 0.6$ and $g_{1} \approx 0.38$ for $f_{ext} = 0.5$.

Figure~\ref{fig:fig3} shows that the model dependence of $g_{1}$ on $\alpha_{pc}$ reproduces, on the whole, the empirical dependence we obtained for a sample of Galactic GCs (Figure~\ref{fig:fig2}), both in terms of the slope and in term of the absolute values. It is evident that the nature of the empirical dependence is more complex as compared with the simplistic nature of the toy model presented here for a particular case of the IMF. In addition to the subsequent dynamical evolution of the GCs, the ($N_{1}(G1)/N_{tot}$)$-\alpha_{pd}$ relation (or equivalently the $g_{1}-\alpha_{pc}$ relation) could potentially be affected by primordial cluster-to-cluster variations of some of the clusters properties. These variations could be either directly measurable ones (e.g., mass, metallicity) or be more subtle such as variations in the IMF of the G1 population, initial mass segregation rations, and binary fractions. We will explore the effects of some of these variations in future work.

Finally, we would like to stress, in other terms, the following important detail we have already noted in the paper text. MS G1 stars from the mass range $[0.1,0.5] M_{\sun}$ are the largest reservoir of these stars, which can merge/collide and form G2 stars in the target mass range (and in the source mass range, too). However, the resulting $g_{1}$ parameter we estimate in the target mass range, $[0.5,0.9] M_{\sun}$, can be smaller (i.e., a smaller fraction of G1 stars in this mass range) at a smaller number of G1 stars consumed in the source mass range. It is due to an additional important factor favoring a smaller proportion between the fraction of G1 and G2 stars in the target mass range. It is the  subtraction  of G1 stars from the target mass range, $[0.5,0.9] M_{\sun}$, for the formation of G2 stars (especially if these G2 stars will be left within this mass range). For example, $0.6 M_{\sun} + 0.2 M_{\sun} = 0.8 M_{\sun}$, i.e. a $0.6 M_{\sun}$ G1 star will be subtracted from the target mass range and consumed for the formation of a $0.8 M_{\sun}$ G2 star in the same mass range.

\subsection{The scenario's implication: a verifiable effect} \label{verif_eff}

The collision-based scenario for the origin of G2 stars described above and which is able to explain the trend in the ($N_{1}(G1)/N_{tot}$)$-\alpha_{pd}$ relation has a number of other implications which can be verified (or tested). Perhaps the most direct would be evidence of the capability of the merging of low-mass stars (at least with $M_{MS} \sim 0.4 M_{\sun}$) to produce, in the collision products, the elemental abundances compatible to those observed in G2 stars. At present, there is unfortunately no definite information on this subject and there is an uncertainty in the understanding of the processes occurring in merging low-mass stars. The characteristics of the finally formed merger, including its surface elemental abundances, are poorly known. Below, we illustrate this with recently obtained relevant observational results (effects), unexpected earlier.

We draw attention to the so-called luminous red novae (LRNe), a class of optical transients. Below, we refer to a particular case of LRNe, which can be the most relevant concerning low-mass mergers in GCs. On the whole, these red transients of different maximum luminosity were interpreted as due either to the merging of the companions of close binary systems \citep[e.g.,][]{sokertylenda2003,kulkaretal2007} or to common envelope evolution of binary systems  \citep[e.g.,][]{blagorodetal2017}. Observational study of LRNe and of their remnants have already brought more understanding of the nature and characteristics of their most probable progenitors and manifested elemental abundances. It has been established that the progenitors are binary systems not only of high and intermediate mass. Among the known LRNe detected to date, there is a low-mass counterpart, V1309 Sco, the most relevant case for our scenario. Its outburst, with maximum luminosity exceeding $10^4 L_{\sun}$, has been caused by the merged stars of a contact binary with the mass of the main companion $\sim 1 M_{\sun}$ \citep{tylendaetal2011} or, more probably, slighter higher, $\sim 1.5 M_{\sun}$. This is now the least massive binary system among those presently known to give rise respectively to the least luminous LRN \citep{howittetal2020}. Recently, \citet{Ferreiraetal2019} studied this merger 10 years after the outburst. They found a dramatic change in its near-IR colour, among other details, and concluded that the the post-outburst state of the merger is consistent with a blue straggler star. \citet{Kaminsketal2018} studied cool molecular outflows produced by three best-known Galactic LRNe, including V1309 Sco, using ALMA observations. Interestingly enough, they found that the molecular composition of these outflows "appear similar to those of oxygen-rich envelopes of classical evolved stars", such as red super-giants or (post-)asymptotic giant branch stars. Although these results are not directly linked to multiple populations, they are linked to our scenario as they clearly indicate that our knowledge of the common-envelope phase of close binary evolution and of its outcome is essentially incomplete. The phenomenon of LRN was not predicted theoretically \citep[e.g.,][]{Goranskijetal2016}. In relation to this, we draw attention to the particularly challenging step of the formation of a merger\slash (collision product) of two MS stars. It is the final formation of its one whole hydrogen burning nucleus from two merging ones. There is significant uncertainty about this process and in particular about the increase in the central temperature as a result of the merger. If the central temperature could temporally achieve at least $\sim(18-20)\times10^6$K or exceed this value, then the CNO-cycle would be activated for some period in a low-mass merger\slash collision product. This is the most important requirement that should be met regarding low-mass merger\slash collision products in order to explain the difference of (at least) their surface CNO abundance from that of G1 stars in the framework of our scenario. Another requirement which has to be met is the availability of a mechanism capable of transporting the products of the CNO-cycle to the external parts of MS G2 stars. The likelihood for such a mechanism to operate is high for low-mass merger\slash collision products. Indeed, they are expected to be fast rotators at least in the early stages after their formation. So, rotational mixing is the likely mechanism that is able to facilitate the transport of synthesized light elements to the stellar surface.

We focused, however, on the scenario's key implication that could be compared with the available observations. We came to the following outcome. Our scenario implies that the parameter $g_{1}$ can vary along the MS. More specifically, as we explained above, it is expected to be, on average, larger (i.e., a higher fraction of G1 stars) in the source mass range than at the upper MS and up to the upper RGB. In particular, it should be true at $M_{MS} < 0.3 M_{\sun}$. In the framework of our scenario of the collisional\slash merging origin of the bulk of G2 stars, we expect a significantly decreased or even a lack of G2 stars (statistically), as compared with their G1 counterparts at $M_{MS} < 0.3 M_{\sun}$. Indeed, if the low mass limit for hydrogen burning (primordial) MS stars is close to $0.1 M_{\sun}$, therefore the low mass limit of collisionally formed G2 stars should strictly speaking be around $0.1 M_{\sun} + 0.1 M_{\sun} = 0.2 M_{\sun}$. It is the lower part of the source mass range that is expected to be least enriched with G2 stars, since the primordial stars capable of forming collision products with such a mass (i.e., $M_{MS} < 0.3 M_{\sun}$) fall within a narrow mass range above the low limit ($M_{MS} \sim 0.1 M_{\sun}$) of the source mass range. It turns out that \citet{milonetal2012,milonetal2014,milonetal2019} estimated the fraction of MS G1 stars below the so-called MS knee (i.e., with $M_{MS} < 0.3 M_{\sun}$) in the GCs NGC 2808, M4, and NGC 6752, respectively. While the fraction of G1 stars on the RGB and the lower MS in NGC 6752 is formally (the authors excluded from consideration the lowest part of the cluster MS where the G1 branch is more extended to a lower luminosity) the same within the error, the situation in two other GCs is strikingly different, especially in NGC 2808. Indeed, in the latter GC, the fraction of G1 stars at the bottom of the MS has been estimated to be around $\sim0.65$ (and characterized from photometry by "primordial helium and enhanced carbon and oxygen abundances"), as compared to the fraction of G1 stars on the RGB around $0.232$ \citep{Milone_etal2017}. In M4 the fractions of G1 stars at the same evolutionary sequences are $\sim0.38$ and $0.285$ \citep{Milone_etal2017}, respectively. These are only three GCs and it is therefore too early to draw any definitive conclusions. However, these results in their actual appearance are rather in agreement with our scenario. A varying proportion between the fractions of G1 and G2 stars with varying stellar mass seems to be less compatible with the canonical way of the formation of G2 stars from a reservoir of gas with a modified elemental abundances appropriate to G2 stars, unless the IMF of the second generation stars is essentially different from that of the primordial generation.

Finally, we note that the most massive GCs with their very small fraction of G1 stars on the RGB, i.e. a low value of the parameter ($N_{1}(G1)/N_{tot}$), around 0.2, seem to be a somewhat special case. One can assume, that the significant contrast between the fraction of G1 stars on the the RGB and on the bottom MS in NGC 2808 is the feature appropriate to the most massive GCs. Partially, it is probably due to the radial effects in GCs we have already mentioned above, namely since the estimates of the parameter ($N_{1}(G1)/N_{tot}$) was done by relying on samples of the centrally located RGB stars in GCs. We cannot exclude that this contrast may additionally depend on metallicity in the sense that it might, on average, be higher at higher metallicity among the most massive GCs. A possible effect of the merging of GCs on the formation of very massive GCs and therefore of their multiple stellar populations should also be kept in mind.

\section{SUMMARY AND CONCLUDING REMARKS} \label{conclusions}

The present paper is devoted to the origin of multiple stellar populations in Galactic GCs. Two generations of stars, G1 and G2, are presently known to typically populate monometallic Galactic GCs. The origin of G2 stars remains an unclear and controversial issue in spite of various plausible astrophysical mechanisms that have been proposed to be responsible for the multiplicity. The current situation is somewhat paradoxical in the sense that thanks to these investigations the origin of generation G2 has become even more mysterious. We report, for the first time, evidence for a potentially crucial role of binary-mediated collisions between G1 low-mass MS stars in the formation of G2 stars in environments of high stellar densities (i.e., the centers of GCs). Such collision products do differ from the primordial counterparts of the same mass. If GCs form with the canonical Kroupa IMF similar to the Galactic field, the characteristic mass of the IMF is around $0.2 M_{\sun}$ and $\sim50$\% of G1 stars have masses in the range $[0.1 - 0.5] M_{\sun}$ (or $\sim85$\% in the range [0.01-0.5] M$_{\sun}$). The IMF evolves with time due to both collisions and dynamical interaction between these low-mass stars. In contrast to the latter, the former process retains these G1 stars in GCs in the form of new stars, i.e. by converting G1 stars into G2 ones of a higher mass. Thereby, it is expected to gradually modify the IMF in the sense: the shallower actual MF of GCs, the smaller is the fraction of G1 stars in the GCs, for a fixed IMF. We explored available data on both the present-day MF and relative number of G1 red giants ($N_{G1}/N_{tot}$) for a sample of 35 GCs. We find that an anti-correlation exists for GCs of high mass ($M_{GC} > 10^{5.3} M_{\sun}$) and is statistically significant at the 98.3\% confidence level. The scatter is larger in this relation for the lest massive GCs ($M_{GC} < 10^{5.1} M_{\sun}$) and the anti-correlation is less significant (i.e., 89\% confidence level).
The importance of the binary-mediated stellar collisions and their prevalence against mass loss by dynamical interactions in high-mass GCs are supported by (a) very significant anti-correlation we found between ($N_{G1}/N_{tot}$) and the stellar encounter rate in GCs and (b) a smaller dynamical disruption of more massive GCs, as predicted by recent simulations.

The populations of both LMXBs and millisecond pulsars in GCs are closely related to the formation of close binaries in the dense central regions of the parent GCs. It seems highly likely that these objects and G2 stars share the same formation mechanism. All these populations in GCs show significant correlation with the stellar encounter rate. We found strong anti-correlation between G1 population and stellar encounter rate.

We note that \citet{Wangetal2020} consider the role of stellar mergers in the formation of multiple populations in GCs with masses lower than $M_{GC} = 3.2\times10^{5.0} M_{\sun}$. However, the authors primarily focus on super- and massive stars and rely on the canonical mode of the formation of G2 stars from the gas ejected by such massive and super-massive stars of population G1. Earlier \citet{Sillsgleb2010} studied the possible contributions of stellar collisions between massive stars to the formation of multiple populations in GCs. They found that such collisions can produce material with elemental abundances comparable to those observed in G2 stars, but the total mass of the material produced is an order of magnitude lower than the required amount. In contrast, our approach advocates for an important role of collision-induced merging of stars belonging to the lower-mass end of the IMF. The collisional formation of G2 MS stars helps to avoid the severe problem of the availability of a sufficient reservoir of gas, since this channel, thanks to a big reservoir of available MS G1 stars, is potentially capable of being the main contributor to the population of G2 stars in GCs. Also, our results point out to a higher efficiency of this mechanism among higher-mass GCs. Of course, a more reliable conclusion about the essential role of collisions between low-mass stars in the formation of G2 stars implies that this mechanism should meet another important requirement. Namely, the ability of the merging of low-mass stars, if only they had the mass of $M_{MS} \sim 0.4 M_{\sun}$, to result in the elemental anti-correlations, such as Na-O, Al-Mg, observed in GC stars from the RGB tip down to the MS turnoff region, i.e. in stars with mass around $0.8 M_{\sun}$. In this regard, we stress the importance of the promising observational results on stellar mergers giving rise to the outbursts observed as luminous red novae, a class of optical transients. Especially interesting and relevant case is a merger of a low-mass contact binary system, with the mass of its main companion $\sim 1 M_{\sun}$, which erupted as a LRN (V1309 Sco). Further work is needed to quantify the relative importance of each of these different mechanisms for the formation of G2 MS stars.

Our results would imply that being negligible or none in low to intermediate density environments, the formation of stars from the merger of other stars can be efficient in regions of high stellar density. Such conditions are appropriate for the central parts of GCs. We estimate that $\sim 1/3$ of the initial number of low-mass stars in a GC should have merged to produce the present-day fraction of G1 stars $\sim0.35$ on the RGB. Hence, we suggest that the collisional nature of G2 stars in GCs can help shed more light on other surprising results obtained for astrophysical objects where stellar encounter rates are high.

From APOGEE data gathered in the Galactic bulge, \citet{Schiavonetal2017} have discovered a population of field stars with high [N/Fe] ratio correlated with [Al/Fe] and anti-correlated with [C/Fe] ratios, that is typical for G2 stars in GCs. However, in contrast with the bimodal metallicity distribution function (MDF) of Galactic GCs, the MDF of these stars is unimodal with its maximum around [Fe/H]$\approx-1.0$ that corresponds to the minimum of the GC MDF in the Milky Way and other galaxies. So, in situ formation of these stars seems to be more probable than their origin from disrupted GCs.

\section{Acknowledgments}

The authors thank the anonymous referee for useful comments that improved the manuscript.
F.A. Calder\'on acknowledges partial support from CONICYT PAI through Grant No. 79170075. J.A. Belinch\'on is supported by CONICYT through FONDECYT Grant No 11170083.

\section*{Data Availability}

The data underlying this article are available in tables from the papers referred to here (at https://doi.org/10.1093/mnras/stw2531; https://doi.org/10.1093/mnras/staa969; https://doi.org/10.1093/mnras/stx1856; https://doi.org/10.1088/0004-637X/766/2/136).

\bibliographystyle{mnras}
\bibliography{paper}

\label{lastpage}

\end{document}